\documentclass[
 reprint,
superscriptaddress,
nofootinbib,
 amsmath,
 amssymb,
 aps,
 nofootinbib
floatfix,
]{revtex4-2}

\usepackage{graphicx}
\usepackage{bm}
\usepackage[colorlinks=true,urlcolor=blue,linkcolor=blue,citecolor=blue,hypertexnames]{hyperref}
\usepackage[nameinlink]{cleveref}
\usepackage{booktabs}
\usepackage{tabularx}
\usepackage{acronym}

\newacro{UV}[UV]{ultraviolet}
\newacro{IR}[IR]{infrared}
\newacro{MS}[MS]{minimal subtraction}
\newacro{DR}[DR]{dimensional regularisation}
\newacro{QFT}[QFT]{quantum field theory}
\newacro{dMS}[dMS]{dimensional minimal subtraction}
\newacro{QCD}[QCD]{quantum chromodynamics}
\newacro{AS}[AS]{asymptotic safety}
\newacro{RG}[RG]{renormalisation group}
\newacro{FRG}[FRG]{functional renormalisation group}
\newacro{PDS}[PDS]{power divergence subtraction}
\newacro{PMS}[PMS]{principle of minimal sensitivity}

\newcommand{\order}[1]{{\cal O}(#1)}

\newcommand{\dderiv}[2]{\frac{\dd #1}{\dd #2}}
\newcommand{\dd}{\text{d}}

  {\list{}{\leftmargin=0.0in\rightmargin=0.0in}  \item[]  }%
  {\endlist}

\crefname{section}{Sec.\!}{Secs.\!}
\crefname{figure}{Fig.\!}{Figs.\!}
\crefname{equation}{}{}
\crefname{table}{Tab.\!}{Tabs.\!}
\crefname{appendix}{App.\!}{Apps.\!}

\begin{document}

\title{Fixed Points of Quantum Gravity from Dimensional Regularisation}

\author{Yannick~Kluth}
\email{yannick.kluth@manchester.ac.uk}
\affiliation{
 Department of Physics and Astronomy, University of Manchester, Manchester M13 9PL, United Kingdom
}

\date{\today}

\begin{abstract}
    We investigate $\beta$-functions of quantum gravity using dimensional regularisation. 
    In contrast to minimal subtraction, a non-minimal renormalisation scheme is employed which is sensitive to power-law divergences from mass terms or dimensionful couplings. 
    By construction, this setup respects global and gauge symmetries, including diffeomorphisms, and allows for systematic extensions to higher loop orders.
    We exemplify this approach in the context of four-dimensional quantum gravity. By computing one-loop $\beta$-functions, we find a non-trivial fixed point. It shows two real critical exponents and is compatible with Weinberg's asymptotic safety scenario. Moreover, the underlying structure of divergences suggests that gravity becomes, effectively, two-dimensional in the ultraviolet.
    We discuss the significance of our results as well as further applications and extensions to higher loop orders.
\end{abstract}
\maketitle

\section{Introduction}
The Einstein-Hilbert action is not renormalisable in four-dimensional spacetime without the inclusion of additional counterterms \cite{Goroff:1985th,Goroff:1985sz,vandeVen:1991gw}. Due to the negative mass dimension of Newton's coupling, each order in a perturbative loop expansion generates \ac{UV} divergences which cannot be subtracted by a renormalisation of the Einstein-Hilbert couplings alone. Within the traditional approach to \ac{QFT}, this means that the theory can only be formulated as an effective field theory below the Planck scale \cite{Donoghue:1994dn}. At energies above the Planck scale, predictivity breaks down since an infinite number of couplings needs to be measured by experiments.

One approach towards a predictive theory of quantum gravity beyond the Planck scale is asymptotic safety \cite{Weinberg:1976xy,Parisi:1977uz,Weinberg:1980gg}. It stipulates the existence of a non-trivial fixed point of the renormalisation group flow. If its number of relevant directions is finite, the theory can be predictive even though an infinite number of operators has to be included in the gravitational action. This mechanism is sometimes also referred to as ``non-perturbative'' renormalisability. Much evidence for the existence of a non-trivial fixed point in quantum gravity has been found using the \ac{FRG} \cite{Wetterich:1992yh,Reuter:1996cp}. For recent reviews, see \cite{Knorr:2022dsx,Morris:2022btf,Platania:2023srt,Martini:2022sll,Wetterich:2022ncl,Eichhorn:2022gku,Saueressig:2023irs,Pawlowski:2023gym}. Notably, these non-perturbative studies frequently observe near-perturbative properties \cite{Eichhorn:2018akn,Eichhorn:2018ydy,Baldazzi:2023pep}, such as the Gaussian scaling of eigenvalues \cite{Falls:2014tra,Falls:2017lst,Falls:2018ylp,Kluth:2020bdv}. 
This suggests that this fixed point may also be found within perturbation theory. In fact, dimensional continuation in \(d = 2 + \varepsilon\) at one loop has provided some of the earliest evidence for the existence of a non-trivial fixed point in four-dimensional quantum gravity \cite{Kawai:1989yh,Jack:1990ey,Kawai:1992np,Kawai:1993fq,Kawai:1993mb,Kawai:1995ju,Falls:2015qga,Martini:2021slj}. 
However, this procedure faces subtleties due to the non-trivial pole structure of the graviton propagator in $d = 2$ \cite{Martini:2022sll}.

As an alternative to dimensional continuation in $d = 2 + \varepsilon$, perturbative quantum gravity can also be studied in $d = 4$. However, results for fixed points seem to depend non-trivially on the chosen renormalisation scheme.
The most commonly used renormalisation scheme in perturbation theory is \ac{DR} with \ac{MS} \cite{tHooft:1972tcz}. Its ability to preserve symmetries, such as gauge invariance or diffeomorphisms, and its straightforward applicability to high order loop calculations \cite{Schnetz:2016fhy,Herzog:2017ohr}, have established \ac{DR} as a central tool to study \ac{QFT}. However, when \ac{DR} with \ac{MS} is applied to quantum gravity in $d = 4 - 2 \varepsilon$, a one-loop fixed point is absent.

The absence of a fixed point in \ac{DR} with \ac{MS} is attributed to the fact that (scheme dependent) power-law divergences are required to observe the fixed point in perturbation theory. In \ac{DR} with \ac{MS}, power-law divergences do not contribute to \(\beta\)-functions.
Nevertheless, by employing renormalisation schemes that track power-law divergences, the non-trivial fixed point can be seen at one loop. Examples include momentum cutoffs, the background covariant operator regularisation, or one-loop approximations of the \ac{FRG} \cite{Liao:1994fp,Codello:2006in,Niedermaier:2009zz,Niedermaier:2010zz}. However, these regularisations present difficulties by breaking the underlying diffeomorphism symmetry \cite{Ebert:2007gf,Pawlowski:2023gym} or do not have appropriate generalisations at higher loop orders \cite{Bilal:2013iva}.

In this letter, we aim at studying quantum gravity within a perturbative renormalisation scheme that retains power-law divergences, and can be straightforwardly generalised to higher loop orders. For this, we employ the non-minimal \ac{PDS} scheme \cite{Weinberg:1980gg,Kaplan:1998tg}. Even though this scheme is based on \ac{DR}, it allows tracking the effects of power-law divergences in perturbative $\beta$-functions. This is done by requiring to subtract all divergences in $d \leq 4$.
We exemplify the usage of \ac{PDS} at the case of one-loop quantum gravity.

\section{Beyond Minimal Subtraction}
\label{sec:cutoff}
In this section, we want to review some basic implications of scheme transformations for the behaviour of $\beta$-functions and their properties. This is followed by a brief discussion of implications for perturbative studies of quantum gravity.

Consider a theory with $\beta$-functions $\beta_i$ and their corresponding couplings $g_i$. A scheme transformation can be represented by a map from the couplings $g_i$ to new couplings $\overline{g}_i$,
\begin{equation}
    \overline{g}_i = g_i \, f_i (g_j) \, .
    \label{eqn:pschemetrafo}
\end{equation}
Since we will focus on perturbative renormalisation schemes below, we choose \cref{eqn:pschemetrafo} such that it does not affect the limit of vanishing couplings.
Using the chain rule, we can derive how $\beta$-functions transform,
\begin{equation}
    \overline{\beta}_i = \dderiv{\overline{g}_i}{\log \mu} = \dderiv{\overline{g}_i}{g_j} \dderiv{g_j}{\log \mu}  = \dderiv{\overline{g}_i}{g_j} \beta_j \, .
    \label{eqn:schemetraf}
\end{equation}
As this expression shows explicitly, $\beta$-functions are not invariant under scheme transformations. However, \cref{eqn:schemetraf} can be used to show the scheme invariance of various properties of $\beta$-functions. It is readily confirmed that zeros (infinities) of $\beta_i$ also imply zeros (infinities) for $\overline{\beta}_i$. This is true as long as scheme transformations are viable, i.e. $\dd \overline{g}_i / \dd g_j$ is finite. In this sense, the existence of fixed points or divergences is scheme invariant.\footnote{Note that the coordinates of zeros or infinities of $\beta$-functions generally change under scheme transformations.} Using \cref{eqn:schemetraf}, it can also be shown that further quantities, such as critical exponents, are invariant.
However, these invariance properties generally only hold for exact expressions. When approximations are used, they can be much more delicate. 

In the case of perturbation theory, issues arise from the fact that \cref{eqn:schemetraf} mixes different orders of the loop expansion. Therefore, \cref{eqn:schemetraf} can only be explicitly verified when working non-perturbatively in the coupling. Thus, the invariance of critical exponents or the existence of fixed points may break down in perturbation theory. This does not immediately render perturbation theory useless to study non-trivial fixed points, but it emphasizes the importance of choosing an appropriate renormalisation scheme. In particular, we may expect that a fixed point, even if it exists in the physical theory, may converge in some, but not all perturbative renormalisation schemes. If the radius of convergence of a perturbative $\beta$-function is finite, this can be explicitly seen by the fact that scheme transformations change the radius of convergence.\footnote{Let us note that questions about the radius of convergence of perturbative $\beta$-functions are difficult to answer in general. On the one hand, one might expect them to be asymptotic, just as observables in \ac{QFT} are expected to be asymptotic in perturbation theory \cite{Dyson:1952tj}. On the other hand, there are explicit results for $\beta$-functions in the literature featuring a perturbative expansion with a non-zero radius of convergence \cite{Novikov:1983uc,Palanques-Mestre:1983ogz,Cresswell-Hogg:2022lgg}. The generality of such statements is further complicated by the possibility of performing scheme transformations.}

This point is also relevant for perturbative studies of the fixed point structure in quantum gravity. While a fixed point can be identified at one loop in schemes that retain power-law divergences \cite{Liao:1994fp,Codello:2006in,Niedermaier:2009zz,Niedermaier:2010zz}, it does not appear in schemes where these divergences do not contribute to $\beta$-functions, such as \ac{DR} with \ac{MS} in $d = 4 - 2 \varepsilon$. Nevertheless, as the discussion above shows, this does not disprove the existence of a non-trivial gravitational fixed point. Instead, we must be aware that the convergence of such a fixed point cannot be guaranteed in any perturbative renormalisation scheme. 
Conversely, if a fixed point can be found in a given renormalisation scheme, it is necessary to observe its convergence towards higher orders before we can conclude that it is a genuine fixed point of the theory.

If power-law divergences are present, we may doubt that schemes like \ac{MS} which simply ignore their effects on $\beta$-functions always lead to the best perturbative convergence. Instead, non-minimal renormalisation schemes may improve convergence by tracking theory information encoded in power-law divergences.
For a perturbative investigation of the fixed point structure in quantum gravity, this suggests that we should employ such non-minimal renormalisation schemes as well.
However, applying them to higher loop orders often results in significant challenges. Momentum cutoff regularisations, or loop expansions of the \ac{FRG}, break diffeomorphism invariance and require the introduction of counterterms to restore the symmetry \cite{Ebert:2007gf,Pawlowski:2023gym}. Schwinger proper time cutoffs, or more generally the background covariant operator regularisation \cite{Liao:1994fp}, do not violate symmetries in combination with the background field method \cite{DeWitt:1967ub,Abbott:1980hw}, but lack a higher loop generalisation \cite{Bilal:2013iva}. Finally, dimensional continuation in $d = 2 + \varepsilon$ is plagued by subtleties due to poles of the graviton propagator in $d = 2$ \cite{Martini:2022sll}. These issues highlight the motivation for focusing on a \ac{DR}-based renormalisation scheme in $d = 4 - 2 \varepsilon$ that is sensitive to power-law divergences. 
Such a scheme naturally respects symmetries such as diffeomorphism invariance, and can be applied straightforwardly to higher loop orders, thus, enabling perturbative explorations of non-trivial fixed points in quantum gravity.

\section{Power Divergence Subtraction}
\label{sec:dimreg}

To understand the behaviour of power-law divergences in more detail, let us start by investigating them in \ac{MS} with a cutoff regulator. For simplicity, we focus on the renormalisation of Newton's coupling $G$ in one-loop quantum gravity. Regularising the theory with an \ac{UV} cutoff $k$, the one-loop renormalisation of $G$ can be given in \ac{MS} by
\begin{equation}
    G_0 = G  + G^2 \left[B_1 k^2 + B_2 \Lambda \log \left( \frac{k}{\mu} \right) \right] \, ,
    \label{eqn:cutoffms}
\end{equation}
with $G_0$ the bare Newton coupling, $\mu$ the sliding scale, and $G$ and $\Lambda$ the renormalised Newton coupling and cosmological constant, respectively. The coefficients $B_i$ are numbers which can be determined from the structure of one-loop divergences. Note that \cref{eqn:cutoffms} does not imply any effects for the $\beta$-function from the power-law divergence. Indeed, using $\tfrac{\dd}{\dd \log \mu} G_0 = 0$, we find
\begin{equation}
    \mu \dderiv{g}{\mu} = 2 g + B_2 g^2 \lambda^2 \, ,
\end{equation}
with $g = \mu^2 G$ and $\lambda = m^{-2} \Lambda$ the dimensionless Newton coupling and cosmological constant, respectively.
Only the logarithmic divergence $B_2$ has left a mark in this $\beta$-function, while the power-law divergence related to $B_1$ is set to zero. This is because power-law divergences in bare couplings do not have an explicit $\mu$-dependence in \ac{MS}. Thus, they do not leave any imprints on the $\beta$-functions. In this way, \ac{MS} sets all contributions of power-law divergences in $\beta$-functions to zero, even if the regulator shows an explicit power-law divergence.

Non-trivial contributions from power-law divergences in \(\beta\)-functions can be obtained using a non-minimal subtraction. An example is given by the Wilson inspired scheme used by Niedermaier at one-loop quantum gravity \cite{Niedermaier:2009zz,Niedermaier:2010zz}. In this scheme, we require that renormalised and bare couplings are equal when the sliding scale is set to the cutoff scale,
\begin{equation}
    G_0(\mu = k) = G \, .
    \label{eqn:niedermaierscheme}
\end{equation}
To ensure \cref{eqn:niedermaierscheme}, the bare coupling must include an additional finite and $\mu$-dependent contribution,
\begin{equation}
        G_0 = G + G^2 \left[ B_1 \left(k^2 - \mu^2\right) + B_2 \Lambda \log \left( \frac{k}{\mu} \right) \right] \, .
        \label{eqn:cutoffniedermaier}
\end{equation}
These terms lead to a non-trivial contribution of power-law divergences in $\beta$-functions,
\begin{equation}
        \mu \dderiv{g}{\mu} = 2 g + 2 B_1 g^2 + B_2 g^2 \lambda^2 \, .
\end{equation}
Thus, non-minimal renormalisation schemes are required to retain the effects of power-law divergences in $\beta$-functions, even if we regularise the theory using a cutoff regulator.

Let us now define a non-minimal renormalisation scheme based on \ac{DR} which leads to non-trivial power-law divergences.
Translating \cref{eqn:cutoffniedermaier} to \ac{DR}, we want to find a renormalisation scheme that fixes the ansatz
\begin{equation}
    G_0 = G + G^2 \left[\widetilde{B}_1 \mu^{d - 2} + B_2 \Lambda \frac{\mu^{d - 4}}{d - 4} \right] \, .
    \label{eqn:PDSansatz}
\end{equation}
The crucial point is a requiring a non-trivial value for $\widetilde{B}_1$.\footnote{The $\mu$-dependent coefficient in front of $\widetilde{B}_1$ follows from dimensional analysis.} A vanishing $\widetilde{B}_1$ corresponds to \ac{MS}.

We can motivate a subtraction scheme that leads to a non-trivial \cref{eqn:PDSansatz} by analysing how power-law divergences can be identified in \ac{DR}. Consider the integral
\begin{equation}
    \int \frac{\dd^d p}{\pi^{d/2}} \frac{1}{p^2 + m^2} = m^{d - 2} \Gamma \left( 1 - \frac{d}{2} \right) \, ,
    \label{eqn:dimregexample}
\end{equation}
which is quadratically divergent in $d = 4$.
Setting $d = 4 - 2 \varepsilon$, we observe a $\tfrac{1}{\varepsilon}$-pole on the right-hand side of \cref{eqn:dimregexample}. This stems from a logarithmic \ac{UV} divergence of the integral which is encountered by an expansion to first order in $m^2$. In contrast, the quadratic divergence is seen as a $\tfrac{1}{\epsilon}$-pole in $d = 2 - 2 \varepsilon$. 
This is a well-known and general feature of \ac{DR} \cite{Veltman:1980mj}: Logarithmic divergences lead to $\tfrac{1}{\varepsilon}$-poles when expanding in $d = 4 - 2 \varepsilon$, while power-law divergences lead to $\tfrac{1}{\varepsilon}$-poles when expanding in $d = n - 2 \varepsilon$, with $n < d$. Power-law divergences are encoded as divergences in lower dimensions.

This fact can be used to define a non-minimal renormalisation scheme in which power-law divergences generate non-trivial contributions to $\beta$-functions. The one that we follow here is given by:
\begin{align}
    \begin{split}
        &\text{\emph{Minimally renormalise bare couplings such that}} \\
        &\text{\emph{they absorb all \ac{UV} divergences in $d \leq 4$.}}
    \end{split}
    \label{scheme}
\end{align}
This scheme can be thought of as a minimal extension of \ac{MS} which captures power-law divergences.
Such a scheme was suggested by Weinberg \cite{Weinberg:1980gg}, but first applied in the context of nucleon-nucleon interactions in effective field theories under the term power divergence subtraction (PDS) \cite{Kaplan:1998tg}.
The requirement of renormalising divergences in $d < 4$ only leads to additional finite contributions in bare couplings when $d = 4$. As such, it is a valid, non-minimal renormalisation scheme based on \ac{DR}.

Let us discuss the application of \ac{PDS} at the example of renormalising Newton's coupling and the cosmological constant at one-loop.
This requires an ansatz of the form,
\begin{equation}
    \begin{split}
        \Lambda_0 =&\, \Lambda + G \left[A_1 \frac{\mu^{d}}{d} + A_2 \Lambda \frac{\mu^{d - 2}}{d - 2} + A_3 \Lambda^2 \frac{\mu^{d - 4}}{d - 4} \right] \, , \\
        G_0 =&\, G + G^2 \left[B_1 \frac{\mu^{d - 2}}{d - 2} + B_2 \Lambda \frac{\mu^{d - 4}}{d - 4} 
    \right] \, .
    \end{split}
    \label{eqn:bareansatz}
\end{equation}
The ansatz \cref{eqn:bareansatz} contains one term for each possible \ac{UV} divergence that can be encountered at a given loop order. The cosmological constant obtains logarithmic, quadratic, and quartic \ac{UV} divergences. In \ac{DR}, these are related to divergences in $d = 4$, $d = 2$, and $d = 0$, respectively, generating three terms in the one-loop renormalisation of $\Lambda_0$ in \cref{eqn:bareansatz}. Newton's coupling only receives logarithmic and quadratic \ac{UV} divergences at one loop. These are related to divergences in $d = 4$ and in $d = 2$, giving rise to two terms in \cref{eqn:bareansatz}.
Each coefficient in \cref{eqn:bareansatz} can be fixed by computing \ac{UV} divergences in the corresponding dimensions.

Using the ansatz \cref{eqn:bareansatz}, we find for the running couplings,
\begin{equation}
    \begin{split}
        \mu \dderiv{\lambda}{\mu} =&\, -2 \lambda - A_1 g - A_2 g \lambda - A_3 g \lambda^2 \, , \\
        \mu \dderiv{g}{\mu} =&\, (d - 2) g - B_1 g^2 - B_2 g^2 \lambda \, ,
    \end{split}
\end{equation}
where we have used the dimensionless Newton coupling as $g = \mu^{d - 2} G$, and the dimensionless cosmological constant as $\lambda = \mu^{-2} \Lambda$.
Thus, following \ac{PDS} we obtain $\beta$-functions that pick up non-trivial contributions from power-law divergences. Below, we will fix the coefficients $A_i$ and $B_i$ using a one-loop computation and analyse the implications.

Let us make some remarks concerning the generalisation of \ac{PDS} to higher loop orders.
Consider \ac{PDS} at $L$ loop with power-law divergences of degree $N$. These power-law divergences generate $\tfrac{1}{\varepsilon}$-poles when $d = d_\text{crit} - 2 \varepsilon$. We call $d_\text{crit}$ the critical dimension. This critical dimension depends on the degree of divergence, and the loop order.
Using dimensional analysis, it can be shown that 
\begin{equation}
    d_\text{crit} = 4 - \frac{N}{L} \, .
    \label{eqn:crit-dim}
\end{equation}
Note that the critical dimension shifts at each loop order for power-law divergences. Only logarithmic divergences ($N = 0$) are consistently found as divergences in the same dimension, namely at $d = 4$.

In renormalisable theories, the critical dimension approaches $d_\text{crit} \to 4$ at high loop orders. This is because such theories have an upper bound on $N$. In non-renormalisable theories, this is in general not the case.
Considering quantum gravity, the strongest divergences depend on the loop order --- they are power-law divergences of order $N = 2L + 2$. Using this in \cref{eqn:crit-dim}, we find
\begin{equation}
    d_\text{crit}^\text{QG} = 2 - \frac{2}{L} \, . 
    \label{eqn:crit-dimqg}
\end{equation}
Thus, the strongest divergences in quantum gravity approach $d_\text{crit}^\text{QG} \to 2$ at infinite loop order.

A technical complication arising at higher loop orders is the renormalisation of subdivergences. An elegant way to deal with them is the incomplete $R$-operation $R'$ \cite{Vladimirov:1979zm}, which is also necessary for an efficient utilisation of the background field method at higher loop orders \cite{Goroff:1985th,vandeVen:1991gw}. To adapt the $R'$ operation for \ac{PDS}, only minor modifications have to be considered that we review now.

Considering a graph $G$ and all of its subgraphs $G_i$, the $R'$-operation is defined recursively by
\begin{equation}
    \begin{split}
        R' G = G + \sum_{\left\{ G_i \right\}}&\, \left( - \mathcal{K} R' G_1 \right) \cdot \dots \cdot \left( - \mathcal{K} R' G_m \right) \times \\
        & \left( G \, \backslash \left\{ G_1, \dots, G_m \right\} \right) \, .
    \end{split}
\end{equation}
The sum runs over all possible combinations of subgraphs $G_i$ of $G$, and $G \, \backslash \left\{ G_1, \dots, G_m \right\}$ denotes the graph $G$ without the subgraphs $G_i$. In \ac{MS}, the operator $\mathcal{K}$ gives the divergence of its argument in $d = 4 - 2 \varepsilon$,
\begin{equation}
    \mathcal{K}^\text{MS} G = \text{div} \left\{ G \big|_{d = 4 - 2 \varepsilon} \right\} \, ,
\end{equation}
with div defined such that
\begin{equation}
    \text{div} \sum_{n} \varepsilon^n = \sum_{n < 0} \varepsilon^n \, ,
\end{equation}
i.e. it extracts the divergent terms from a Laurent series in $\varepsilon$.
To employ \ac{PDS}, the operator $\mathcal{K}$ needs to be modified. It does not only need to return the divergence in $d = 4$, but divergences in all critical dimensions $d_\text{crit}$. For that purpose, we define,
\begin{equation}
    \mathcal{K} G = \sum_{N = 0}^{N_\text{max}} \text{div} \left\{ G \big|_{d = d_\text{crit} (N, L) - 2 \varepsilon} \right\} \, ,
\end{equation}
where $L$ is the loop order of the diagram $G$, and $N_\text{max}$ the strongest divergence that can be generated by $G$. This allows for $\mathcal{K}$ to pick up all relevant \ac{UV} divergences in $d \leq 4$. In this way, \ac{PDS} can be combined with the $R'$-operation to subtract all subdivergences at higher loop orders. With this in mind, \ac{PDS} can be applied straightforwardly to higher loop orders, in particular also in combination with the background field method in quantum gravity.

\section{Quantum Gravity at One Loop}
\label{sec:oneloop}
In this section, we apply \ac{PDS} to quantum gravity at one loop. The gravitational action that we consider is given by
\begin{equation}
    S = \int \dd^d x \sqrt{g} \, \left[ \frac{\Lambda_0}{8 \pi G_0} - \frac{R}{16 \pi G_0}
    \right] \, .
    \label{eqn:action}
\end{equation}
Using a harmonic gauge fixing in the background field method \cite{Buchbinder:1992rb}, we compute one loop 
divergences from the one-loop effective action $\Gamma^{(1)}$. This computation relies on the graviton and ghost propagators. As described in \cref{app:continuation}, we use a non-trivial procedure to dimensionally continue the graviton propagator away from $d = 4$. This modification allows us to avoid a singular graviton propagator in $d = 2$. Following \cref{app:continuation}, we have
\begin{equation}
    \Gamma^{(1)} = \frac{1}{2} \text{Tr} \log \big\{ \mathcal{H} \big\} - \text{Tr} \log \Big\{ S^{(2)}_\text{gh} \Big\} \, ,
    \label{eqn:oneloopeffaction}
\end{equation}
where $\mathcal{H}$ is the dimensionally continued version of the graviton Hessian fulfilling
\begin{equation}
    \mathcal{H} \Big|_{d = 4} = S^{(2)} + S_\text{gf}^{(2)} \Big|_{d = 4} \, .
\end{equation}
For an explicit definition of $\mathcal{H}$, see \cref{eqn:hessianproperdef}.

Following \ac{PDS} and the structure of one-loop \ac{UV} divergences encoded in the ansatz \cref{eqn:bareansatz}, we compute the divergences of \cref{eqn:oneloopeffaction} in $d = 4$, $d = 2$, and $d = 0$.
The one-loop divergences in $d = 4 - 2 \varepsilon$ are given by
\begin{equation}
    \begin{split}
        \Gamma^{(1)} \Big|_{d = 4 - 2 \varepsilon} = &\,\frac{1}{(4 \pi)^2 \varepsilon} \bigg[ - 10 \Lambda_0^2 + \frac{13}{3} \Lambda_0 R - \frac{53}{90} E \\
        & - \frac{7}{20} R_{\mu \nu} R^{\mu \nu} - \frac{1}{120} R^2 \bigg] + \order{1} \, ,
    \end{split}
    \label{eqn:oneloop4}
\end{equation}
with $E = R_{\mu \nu \rho \sigma} R^{\mu \nu \rho \sigma} - 4 R_{\mu \nu} R^{\mu \nu} + R^2$ the topological Euler density.
These are the (scheme independent) logarithmic divergences. 

The quadratic divergences computed in $d = 2 - 2 \varepsilon$ take the form,
\begin{equation}
        \Gamma^{(1)} \Big|_{d = 2 - 2 \varepsilon} = \frac{1}{(4 \pi) \varepsilon} \left[ -3 \Lambda_0 + \frac{31}{12} R \right] + \order{1} \, .
        \label{eqn:oneloop2}
\end{equation}
Contrary to the logarithmic divergences, these are scheme dependent. In particular, specific choices of our scheme, such as the dimensional continuation of the propagator described in \cref{app:continuation} affect the result in \cref{eqn:oneloop2}. 

In $d = - 2 \varepsilon$ we do not find any divergences,
\begin{equation}
    \Gamma^{(1)} \Big|_{d = - 2 \varepsilon} = \order{1} \, .
    \label{eqn:oneloop0}
\end{equation}
This can be understood by writing the traces in \cref{eqn:oneloopeffaction} in terms of heat kernels.
Denoting the Seeley-DeWitt coefficients that enter the traces by $A_i$, the divergences in $d = 0$ are captured by $A_0$, see also \cref{eqn:heatkernel}. The trace of $A_0$, which enters the one-loop effective action \cref{eqn:oneloopeffaction}, is given by $\text{Tr} \left\{ A_0 \right\} = \frac{d (d + 1)}{2}$ for the graviton, and by $\text{Tr} \left\{ A_0 \right\} = d$ for the ghost. Note that both are proportional to a factor of $d$. It is this factor that cancels the \ac{UV} divergence encountered in $d = 0$. While the \ac{UV} divergence results in a pole of the form of $\tfrac{1}{d}$ from a Schwinger integral \cref{eqn:heatkernel}, the trace of $A_0$ cancels the factor of $d$ in the denominator, leading to a finite result. 

Note that \cref{eqn:oneloop0} is not affected by altering the dimensional continuation of the Hessian $\mathcal{H}$. Since $A_0$ is given by the identity element of the field space, it is independent of any endomorphisms included in $\mathcal{H}$. Moreover, prefactors in front of the differential operator in $\mathcal{H}$ drop out due to the logarithm in \cref{eqn:oneloopeffaction}. Unless we define a dimensional continuation of the propagator which completely alters the form of the differential operator, such as making it non-minimal, the result \cref{eqn:oneloop0} seems to be hard-wired in \ac{PDS}. In this context, it is interesting to note that similar conclusions about the absence of quartic divergences have been made in \cite{Akhmedov:2002ts,Ossola:2003ku}.

We now use the results \cref{eqn:oneloop4,eqn:oneloop2,eqn:oneloop0} to determine the coefficients in our ansatz for the bare couplings \cref{eqn:bareansatz}.
Since the Ricci tensor vanishes on-shell, and $E$ is a topological invariant, we will discard the divergences related to quadratic curvature invariants here. They could be taken into account, for example, using field redefinitions \cite{Baldazzi:2021ydj,Baldazzi:2021orb}.
However, for brevity, we will simply neglect these contributions instead.\footnote{Following from the vanishing of quartic divergences in \cref{eqn:oneloop0}, one can show that the minimal essential scheme of \cite{Baldazzi:2021orb} simply corresponds to keeping $\lambda = 0$ in \cref{eqn:betafunctions}. In particular, the divergences related to quadratic curvatures do not affect the $\beta$-function for Newton's coupling.}
We find
\begin{equation}
    \begin{split}
        \Lambda_0 =&\, \Lambda  + G \left[\frac{26}{3} \Lambda \frac{\mu^{d - 2}}{d - 2} - \frac{4}{3 \pi} \Lambda^2 \frac{\mu^{d - 4}}{d - 4} \right] \, , \\
        G_0 =&\, G + G^2 \left[\frac{62}{3} \frac{\mu^{d - 2}}{d - 2} + \frac{26}{3 \pi} \Lambda \frac{\mu^{d - 4}}{d - 4} \right] \, .
    \end{split}
\end{equation}
This gives rise to the $\beta$-functions
\begin{equation}
    \begin{aligned}
        &\beta_\lambda =&\, \mu \dderiv{\lambda}{\mu} =&\, -2 \lambda - \frac{26}{3} g \lambda + \frac{4}{3 \pi} g \lambda^2 \, , \\
        &\beta_g =&\, \mu \dderiv{g}{\mu} =&\, 2 g - \frac{62}{3} g^2 - \frac{26}{3 \pi} g^2 \lambda \, .
    \end{aligned}
    \label{eqn:betafunctions}
\end{equation}
It is worth noting that $\beta_\lambda = 0$ for $\lambda = 0$, independently of the value of Newton's coupling $g$. This originates from the vanishing of quartic divergences, see \cref{eqn:oneloop0}. If the quartic divergences were non-trivial, there would be an additional contribution to $\beta_\lambda$ of the form of $a g$, with $a$ a number. In that case, the running of Newton's coupling would immediately imply a running for the cosmological constant, and it would be impossible to keep $\lambda = 0$ fixed for any value of $g$.

\begin{table}
    \begin{tabularx}{\columnwidth}{l c c >{\centering\arraybackslash}X >{\centering\arraybackslash}X}
        \toprule
         & \(\lambda\) & \(g\) & \(\theta_0\) & \(\theta_1\) \\ 
        \midrule
        \(\textbf{FP}_\text{Gauss}\) & \(0\) & \(0\) & \(2\) & \(-2\) \\ 
        \midrule
        \(\textbf{FP}_\text{UV}\) & \(0\) & \(\frac{3}{31}\) & \(2.839\) & \(2\) \\ 
        \midrule
        \(\textbf{FP}_\text{unphys}\) & \(-4 \pi\) & \(-\frac{1}{7}\) & \(0.619 + 2.828i\) & \(0.619 - 2.828i\) \\ 
        \bottomrule
    \end{tabularx}
    \caption{Fixed point values and their critical exponents for the $\beta$-functions found with \ac{PDS} at one loop in quantum gravity. }
    \label{tab:fixedpoints}
\end{table}

The $\beta$-functions \cref{eqn:betafunctions} feature a total of three fixed points shown in \cref{tab:fixedpoints}. This includes the Gaussian fixed point $\textbf{FP}_\text{Gauss}$ located at vanishing couplings, a non-trivial \ac{UV} fixed point $\textbf{FP}_\text{UV}$, and an unphysical fixed point $\textbf{FP}_\text{unphys}$ located at negative Newton coupling. The fixed point $\textbf{FP}_\text{UV}$ features a vanishing cosmological constant, and a positive Newton coupling of $g \approx 0.0968$. The fact that $\lambda = 0$ is closely related to the vanishing of the quartic divergence in \cref{eqn:oneloop0}. Due to that, $\lambda = 0$ is a fixed point of $\beta_\lambda$ and the fixed point value of $g$ only has to lead to a vanishing of $\beta_g$.

Both critical exponents of $\textbf{FP}_\text{UV}$ are positive, meaning that the fixed point has two relevant directions. They are also real, with the cosmological constant giving a critical exponent of $\theta_0 \approx 2.839$, while the critical exponent related to Newton's coupling is given by $\theta_1 = 2$. Note that $\theta_1$ is fixed by the fact that we perform a one-loop computation and observe a vanishing cosmological constant at the fixed point. This dictates $\theta_1 = 2$. Compared to the literature, both values for the critical exponents fall within expected results from non-perturbative computations. However, note that these critical exponents are usually seen as complex conjugate pairs. For a discussion on this topic, see \cite{Falls:2014tra}. A disentanglement into two real exponents is sometimes observed in cases where the cosmological constant decouples in some way \cite{Ohta:2015efa,Ohta:2015fcu,Baldazzi:2021orb,Kluth:2022vnq}. In our case, a similar effect takes place in the form of a vanishing quartic divergence for the cosmological constant in \cref{eqn:oneloop0}.

\begin{figure}
    \includegraphics[width=\columnwidth]{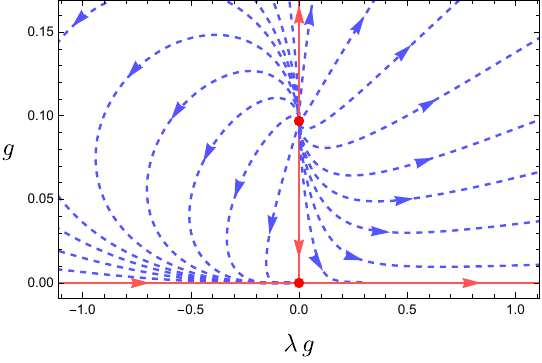}
    \caption{The one-loop phase diagram of quantum gravity showing Newton's coupling $g$ and the product of Newton's coupling and the cosmological constant $g \lambda$. Red dots indicate the non-trivial fixed point $\textbf{FP}_\text{UV}$ and the Gaussian fixed point $\textbf{FP}_\text{Gauss}$. Blue dashed line illustrate trajectories with arrows pointing to the IR. Separatrices are shown as red solid lines.}
    \label{fig:phase}
\end{figure}

In \cref{fig:phase} we show the phase diagram of the $\beta$-functions in \cref{eqn:betafunctions} in the physical region $g > 0$. Note that we have translated the couplings $\{\lambda, g\}$ into the set $\{\lambda g, g\}$. As discussed above, the non-trivial fixed point $\textbf{FP}_\text{UV}$ has two relevant directions. Thus, all trajectories in its vicinity end up in $\textbf{FP}_\text{UV}$ in the \ac{UV}. Trajectories flowing out of $\textbf{FP}_\text{UV}$ can end up in two different regions. In case they flow towards negative cosmological constant, they show some intermediate scaling, but end up at $\lambda g = g = 0$ in the IR. This does not imply that the cosmological constant vanishes in the IR for these trajectories. In fact, the only trajectory connecting $\textbf{FP}_\text{UV}$ to a vanishing cosmological constant in the IR is the separatrix on which the cosmological constant vanishes all along. All other trajectories flowing to $\lambda g = 0$ in the IR actually have a diverging (negative) cosmological constant in the IR. Trajectories flowing out of $\textbf{FP}_\text{UV}$ towards positive cosmological constants lead to $\lambda g$ growing towards large non-perturbative values in the IR.

Note that the existence of $\textbf{FP}_\text{UV}$ implies a cancellation between the one-loop and the tree-level contributions to $\beta_g$. As such, care must be taken when interpreting its physical significance. However, the fact that its properties resemble non-perturbative results indicate that it might be related to a true physical fixed point of quantum gravity. Moreover, the smallness of Newton's coupling suggests that the fixed point might be perturbative enough to converge once higher order loop corrections are taken into account. However, it should be noted that this argument is only valid if the coefficients of higher order loop corrections in the $\beta$-functions do not grow too rapidly. With this in mind, a computation of higher loop coefficients with \ac{PDS} would be of great interest. 

Finally, let us point out that $\textbf{FP}_\text{UV}$ is induced entirely by the \ac{UV} divergence related to $B_1$ in \cref{eqn:bareansatz}. This is due to the fact that the cosmological constant vanishes. As a consequence, the UV behaviour of $\textbf{FP}_\text{UV}$ is solely determined by divergences encountered in $d = 2$. 

\section{Conclusions}
\label{sec:conclusion}

In this letter, we have employed dimensional regularisation without minimal subtraction, such that we can account for non-trivial power-law divergences. Such divergences arise naturally in theories with dimensionful couplings, in particular, perturbatively non-renormalisable theories. We have argued that retaining power-law divergences can be beneficial to capture key properties of such theories.
For any $d$-dimensional theory, this can be done by subtracting all divergences in dimensions $\leq d$. This scheme is called \ac{PDS} \cite{Weinberg:1980gg,Kaplan:1998tg} and preserves underlying global or gauge symmetries, as well as diffeomorphisms. Moreover, it is conceivable that this procedure improves the convergence of perturbative expansions.

We have applied our setup to four-dimensional quantum gravity at one loop. The resulting $\beta$-functions in \cref{eqn:betafunctions} give rise to a non-trivial fixed point suitable for the asymptotic safety scenario, see \cref{tab:fixedpoints}. Moreover, its properties are consistent with results obtained from cutoff regularisations \cite{Reuter:1996cp,Codello:2006in,Niedermaier:2009zz,Niedermaier:2010zz,Saueressig:2023irs,Pawlowski:2023gym,Morris:2022btf}. While such schemes usually face difficulties at higher loop orders \cite{Bilal:2013iva} or break underlying symmetries of the theory \cite{Ebert:2007gf,Pawlowski:2023gym}, our setup is free from such subtleties.
Furthermore, it is noteworthy that our critical exponents are real rather than complex conjugate pairs \cite{Falls:2014tra}.

An interesting feature of our results is the impact played by \ac{UV} divergences in $d = 2$. Using \cref{eqn:crit-dimqg}, we have found that the strongest divergences in quantum gravity approach $d \to 2$ in the infinite loop limit. In this sense, the strongest \ac{UV} divergences are associated with two dimensional spacetime.
Moreover, in the one-loop computation we have found quartic divergences to be absent. As a consequence, the cosmological constant vanishes at the fixed point and the fixed point is entirely determined by divergences in $d = 2$. These observations could be interpreted as ramifications that quantum gravity becomes two-dimensional in the \ac{UV}. This conjecture has been observed across several different approaches to quantum gravity \cite{Carlip:2017eud}.

Our results for one-loop quantum gravity ask for extensions to higher loop orders, particularly to assess the convergence of the non-trivial fixed point. The smallness of Newton's coupling suggests that this might be possible. A first step towards this is the computation of quantum gravity at two loop. With a recent computation suggesting that the effects of the two-loop counterterms may be only marginal \cite{Baldazzi:2023pep}, we could expect convergence of our results as well. The two-loop computation would also give rise to a non-trivial critical exponent for Newton's coupling for the first time. This could be compared to non-perturbative results and provide another benchmark test for the use of \ac{PDS} in quantum gravity.

More generally, we note that applications of \ac{PDS} have been focussed on effective field theories, in particular, nuclear interactions \cite{Kaplan:1998we,Bedaque:2002mn,Epelbaum:2008ga,Brivio:2017vri}.
However, as we have shown here, the effects of power-law divergences may also be relevant for other non-renormalisable field theories. Apart from pure quantum gravity, examples include the non-linear sigma model \cite{Brezin:1975sq,Codello:2008qq,Percacci:2009fh,Efremov:2021fub}, four-fermi theories \cite{Nambu:1961tp,Nambu:1961fr,Gross:1974jv,Rosenstein:1988pt,Cresswell-Hogg:2022lgg}, non-abelian gauge theories in $d > 4$ \cite{Morris:2004mg}, or gravity-matter systems \cite{Ebert:2007gf,Rodigast:2009zj,Eichhorn:2022gku}. In the context of gravity, the perturbative approach presented here could also be useful to study the scattering of gravitons within asymptotic safety \cite{Draper:2020bop,Draper:2020knh,Knorr:2022dsx}.

On a different tack, it would also be valuable to explore the implications of other non-minimal renormalisation schemes. 
Since \ac{PDS} is just one out of many schemes of dimensional regularisation that retains the effects of power-law divergences, we could consider modifications to enhance perturbative convergence. This could be achieved using independent optimisation criteria, such as the principle of minimal sensitivity \cite{Stevenson:1981vj}. Better understood theories, such as four-fermi theories could act as valuable toy models to test such ideas.

\section*{Acknowledgments}
    I would like to thank Gabriel Assant, Kevin Falls, Daniel Litim, and Peter Millington for helpful discussions and Daniel Litim and Peter Millington for comments on earlier versions of this manuscript. This work was supported by a United Kingdom Research and Innovation (UKRI) Future Leaders Fellowship [Grant No.~MR/V021974/2].

\section*{Data Access Statement}

No new data were created or analysed in this study.

\appendix

\section{Inversion of Propagator}
\label{app:continuation}
In this section, we describe a modified dimensional continuation of the graviton propagator given by the Hessian $\mathcal{H}$ that has been used in \cref{eqn:oneloopeffaction}. This modification is necessary to avoid a singular propagator in $d = 2$ that would follow from the conventional dimensional continuation. 

We start with the Hessian of the gravitational action,
\begin{equation}
    \overline{\mathcal{H}} = S^{(2)} + S_\text{gf}^{(2)} \, .
\end{equation}
Following our gauge choice, this results in
\begin{equation}
    \begin{split}
        \overline{\mathcal{H}}^{\mu \nu}_{\ \ \rho \sigma} = \frac{1}{32 \pi G_0} \Big[&\, \overline{K}^{\mu \nu}_{\ \ \rho \sigma} \left(- \nabla^2 - 2 L + R\right) \\
        & + R^{\mu \nu} g_{\rho \sigma} + g^{\mu \nu} R_{\rho \sigma} \\
        & - 2R^{\mu \ \nu}_{\ \rho \ \sigma} - 2 R^{(\mu}_{\ \ (\rho} g^{\nu)}_{\ \ \sigma )} \Big] \, ,
    \end{split}
\end{equation}
with
\begin{equation}
    \overline{K}^{\mu \nu}_{\ \ \rho \sigma} = g^\mu_{\ ( \rho} g^\nu_{\ \sigma)} - \frac{1}{2} g^{\mu \nu} g_{\rho \sigma} \, .
\end{equation}
The inverse of $\overline{\mathcal{H}}$ does not exist in $d=2$ due to poles arising from the inverse of $\overline{K}$,
\begin{equation}
    \left( \overline{K}^{-1} \right)^{\mu \nu}_{\ \ \rho \sigma} = g^\mu_{\ ( \rho} g^\nu_{\ \sigma)} + \frac{1}{2 - d} g^{\mu \nu} g_{\rho \sigma} \, .
\end{equation}
Thus, the naive dimensional continuation of the propagator, which is the inverse of $\overline{\mathcal{H}}$, is singular in $d = 2$. This would lead to additional poles unrelated to \ac{UV} divergences when employing \ac{DR} in $d = 2$.

The problem can be circumvented by employing a different dimensional continuation of $\overline{\mathcal{H}}$, leading to a modified form of the propagator in $d = 2$. This is possible since \ac{DR} is solely employed as a regularisation procedure to render expressions finite that would otherwise diverge in $d = 4$. As such, the only requirements on a dimensional continuation of $\overline{\mathcal{H}}$ in dimensions other than $d = 4$ is that it respects symmetries, its propagator is analytic in $d$, and it gives back $\overline{\mathcal{H}}$ in $d = 4$.

The dimensional continuation of the Hessian that has been used in this work is given by
\begin{equation}
    \begin{split}
        \mathcal{H}^{\mu \nu}_{\ \ \rho \sigma} = \frac{1}{32 \pi G_0} \Big[&\, K^{\mu \nu}_{\ \ \rho \sigma} \left(- \nabla^2 - 2 L + R\right) \\
        & + R^{\mu \nu} g_{\rho \sigma} + g^{\mu \nu} R_{\rho \sigma} \\
        &- 2R^{\mu \ \nu}_{\ \rho \ \sigma} - 2 R^{(\mu}_{\ \ (\rho} g^{\nu)}_{\ \ \sigma )} \Big] \, ,
    \end{split}
    \label{eqn:hessianproperdef}
\end{equation}
with 
\begin{equation}
    K^{\mu \nu}_{\ \ \rho \sigma} = g^\mu_{\ ( \rho} g^\nu_{\ \sigma)} + \frac{1}{2 - d} g^{\mu \nu} g_{\rho \sigma} \, .
    \label{eqn:Kmatrix}
\end{equation}
The only difference between $\mathcal{H}$ and $\overline{\mathcal{H}}$ is the replacement of $\overline{K}$ in favour of $K$.
It is straightforward to verify 
\begin{equation}
    \mathcal{H}^{\mu \nu}_{\ \ \rho \sigma} \Big|_{d = 4} = \overline{\mathcal{H}}^{\mu \nu}_{\ \ \rho \sigma} \Big|_{d = 4} \, .
    \label{eqn:hessianproper}
\end{equation}
Even though \cref{eqn:hessianproperdef} with \cref{eqn:Kmatrix} has a pole in $d = 2$, its inverse, i.e. the propagator, is well-defined and analytic in any dimension. To check this explicitly, we can determine the inverse of \cref{eqn:hessianproper} using heat kernel techniques. For this purpose, we factor out the matrix $K$ and write $\mathcal{H}$ in the form
\begin{equation}
    \mathcal{H}^{\mu \nu}_{\ \ \rho \sigma} = \frac{1}{32 \pi G_0} K^{\mu \nu}_{\ \ \alpha \beta} \mathcal{U}^{\alpha \beta}_{\ \ \rho \sigma} \, .
    \label{eqn:hessianfactored}
\end{equation}
The operator $\mathcal{U}$ is a Laplacian of the form
\begin{equation}
    \mathcal{U}^{\mu \nu}_{\ \ \rho \sigma} = - \bm{1}^{\mu \nu}_{\ \ \rho \sigma} \nabla^2 + \bm{E}^{\mu \nu}_{\ \ \rho \sigma} \, ,
\end{equation}
with $\bm{E}$ an endomorphism,
\begin{equation}
    \begin{split}
        \bm{E}^{\mu \nu}_{\ \ \rho \sigma} =&\, \bm{1}^{\mu \nu}_{\ \ \rho \sigma} \left(- 2L + R\right) - 2 R^{\mu \ \nu}_{\ \rho \ \sigma} - \frac{R}{2} g^{\mu \nu} g_{\rho \sigma} \\
        & - 2 R^{(\mu}_{\ \ (\rho} g^{\nu )}_{\ \ \sigma)} + R^{\mu \nu} g_{\rho \sigma} + \frac{6 - d}{2} g^{\mu \nu} R_{\rho \sigma} \, .
    \end{split}
    \label{eqn:endo}
\end{equation}
Note that the endomorphism is only symmetric in $d = 4$. This originates from the properties of $K$ which is its own inverse only in $d =4$. Therefore, the endomorphism $\bm{E}$ and the Laplacian $\mathcal{U}$ are not symmetric in $d \neq 4$. Only in combination with the matrix $K$, the Hessian $\mathcal{H}$ and its inverse are symmetric in any dimension. 

Following from \cref{eqn:hessianfactored}, the inverse of $\mathcal{H}$ is given by
\begin{equation}
    \begin{split}
        \mathcal{P}^{\mu \nu}_{\ \ \rho \sigma} =&\, \left(\mathcal{H}^{-1}\right)^{\mu \nu}_{\ \ \rho \sigma} \\
        =&\, 32 \pi G_0 \left(\mathcal{U}^{-1}\right)^{\mu \nu}_{\ \ \alpha \beta} \left(K^{-1}\right)^{\alpha \beta}_{\ \ \rho \sigma} \, ,
    \end{split}
\end{equation}
with
\begin{equation}
    \left(K^{-1}\right)^{\mu \nu}_{\ \ \rho \sigma} = g^\mu_{\ ( \rho} g^\nu_{\ \sigma)} - \frac{1}{2} g^{\mu \nu} g_{\rho \sigma} \, ,
\end{equation}
well-defined in any dimension.
The inverse of $\mathcal{U}$ is found using \cite{Avramidi:2000bm,Vassilevich:2003xt,Kluth:2019vkg}
\begin{equation}
    \mathcal{U}^{-1} = \int_0^\infty \dd s \, e^{-s \mathcal{U}} = \frac{e^{- \tfrac{\sigma}{2 s}}}{(4 \pi s)^{d/2}} \sum_{n = 0}^\infty s^n A_n  \, ,
    \label{eqn:heatkernelpre}
\end{equation}
with $A_n$ the Seeley-DeWitt coefficients, and $\sigma$ the Synge world function. Since the endomorphism \cref{eqn:endo} is well-defined in any dimension, so are the Seeley-Dewitt coefficients entering \cref{eqn:heatkernelpre}.

The logarithm of $\mathcal{U}$, which is required at one loop, can be given as
\begin{equation}
    \log \frac{\mathcal{U}}{\mathcal{U}_0} = \int_0^\infty \frac{\dd s}{s} \, \left[ e^{-s \mathcal{U}_0} - e^{-s \mathcal{U}} \right] \, ,
\end{equation}
where $\mathcal{U}_0$ is a field independent normalisation, whose contribution vanishes in \ac{DR}. For the graviton trace, this results in
\begin{equation}
    \text{Tr} \log \mathcal{H} = - \sum_{n = 0}^\infty \int_0^{\tfrac{1}{k_\text{IR}^2}} \dd s \, \frac{s^{n - 1}}{(4 \pi s)^{d/2}} \text{Tr} \left\{ A_n \right\} \, .
    \label{eqn:heatkernel}
\end{equation}
Note that we have introduced an IR regulator $k_\text{IR}$ to regularise IR divergences which originate from the Schwinger integral. If we were to calculate physical observables, we would have to take the limit $k_\text{IR} \to 0$. Here, we are only interested in \ac{UV} divergences. Keeping only $\tfrac{1}{\epsilon}$-poles after solving the Schwinger integral with \ac{DR}, the IR regulator drops out.

\bibliography{../../Bib/central-bib.bib}

\end{document}